\documentclass[aps,prd,showpacs,nobibnotes,twocolumn,
nobalancelastpage,superscriptaddress]{revtex4}
\newcommand{\be}{\begin{equation}}
\newcommand{\ee}{\end{equation}}
\usepackage{graphicx}
\usepackage{psfrag}
\usepackage{epsfig}
\usepackage{subfigure}
\begin{document}

\title{GLAST sensitivity to Point Sources of Dark Matter Annihilation}

\author{Gianfranco Bertone} 
\affiliation{INFN, Sezione di Padova, Via Marzolo 8, Padova I-35131, Italy}
\author{Torsten Bringmann}
\affiliation{SISSA/ISAS, Via Beirut 2-4, 34013 Trieste, Italy}
\author{Riccardo Rando}
\affiliation{INFN, Sezione di Padova, Via Marzolo 8, Padova I-35131, Italy}
\author{Giovanni Busetto}
\affiliation{INFN, Sezione di Padova, Via Marzolo 8, Padova I-35131, Italy}
\author{Aldo Morselli}
\affiliation{INFN and University of Roma Tor Vergata, via della Ricerca Scientifica 1, Roma, Italy}

\begin{abstract}
We study the prospects for detecting gamma-rays from point sources of 
Dark Matter annihilation with the space satellite GLAST. We simulate
the instrument response to the gamma-ray spectrum arising from the
annihilation of common Dark Matter candidates, and derive
full-sky sensitivity maps for the {\it detection} of point sources and for
the {\it identification} of the Dark Matter (as opposed to astrophysical)  
origin of the gamma-ray emission. These maps represent a powerful tool to
assess the detectability of point sources, i.e. sources with angular
size smaller than the angular resolution of GLAST, $\Delta \theta\sim 0.1 ^\circ$, 
in {\it any} DM scenario. As
an example, we apply the obtained results to the so-called {\it mini-spikes} 
scenario, where the annihilation signal originates from large 
Dark Matter overdensities around Intermediate Mass 
Black Holes. We find that if these objects exist in the Galaxy, not only GLAST 
should be able to detect them over a timescale as short as 2 months, 
but in many cases it should be possible to determine with good accuracy the mass of
the annihilating Dark Matter particles, while null searches would place
stringent constraints on this scenario. 
\end{abstract}

\maketitle

\section{Introduction}

In the cosmological concordance model that has emerged over the past few years, a considerable fraction of the total energy density today, $\Omega_{CDM}=0.19$, consists of non-baryonic, collisionless and dissipation-free (i.e.~cold) matter \cite{Spergel:2006hy}. To unveil the -- so far completely unknown -- nature of this dark matter (DM) is  one of the most outstanding challenges for cosmology and astroparticle physics today.
The list of proposed DM candidates is long, ranging from modified theories of gravity, that would effectively mimic a large non-relativistic component in the total energy content of the universe, to a whole zoo of speculative new particles that the DM may consist of. While the former approach may, in fact, be extremely successful to describe certain isolated phenomena, like the flattening of galactic rotation curves \cite{MOND}, it is notoriously difficult to reconcile with the whole range of accessible observations; in the following, we will therefore restrict ourselves to the latter possibility (see Refs.~\cite{review1,review2} for recent reviews on particle DM). 

Search strategies for DM particles (going beyond merely testing their gravitational influence) can be grouped into two categories.
In \emph{direct} detection experiments, one tries to trace these particles by looking for the recoil energy they would transfer during scattering events with the atoms of the detector material; null searches start to place non-trivial bounds on the parameter space of the underlying models (see, e.g., \cite{Munoz:2003gx} and references therein). Alternatively, one can use \emph{indirect} detection techniques, making use of the fact that DM particles will generally pair-annihilate in regions of enhanced DM densities; the decay products may then be revealed as exotic contributions to astrophysical fluxes of gamma-ray, neutrino and anti-matter. 

The space satellite GLAST is expected to play a crucial role in indirect DM searches, thanks both to
its ability to perform observations at energy scales comparable to the mass of common
DM candidates and to its potential of making deep full-sky maps in gamma-rays, thanks to its large ($\sim 2.4$ sr) field-of-view~\cite{glast}. 
In the first part of this paper, we study the prospects for detecting point sources of DM annihilation with GLAST, and assess the possibility
to discriminate them from ordinary astrophysical sources. As a result of this 
analysis, we obtain full-sky sensitivity maps for the {\it detection} of sources above the 
diffuse background, as well as for the {\it identification} of DM annihilation sources, i.e. 
for the detection of sources that can be discriminated against ordinary astrophysical sources. 

This analysis can be applied to any astrophysical scenario that predicts the existence of point-like sources of gamma-rays from DM annihilations, where ``point-like'' in this context means sources with an angular size smaller than the GLAST angular resolution, $\Delta \theta \sim 0.1^\circ$ (see below for further details). In the second part of the paper, in order to show the 
effectiveness of this approach, we
apply our results to a specific astrophysical scenario, where gamma-rays originate from DM {\it mini-spikes}. In fact, it has been recently pointed out that intermediate mass black holes (IMBHs) are a particularly promising place to look for DM self-annihilations,  since,  under certain assumptions about their formation mechanism, they will be surrounded by regions of highly enhanced DM density, called mini-spikes \cite{Bertone:2005xz}. As a consequence, they are expected to appear as bright sources of gamma rays, thereby providing a new possible means of indirect DM detection. Using the sensitivity maps obtained in the first part of the present paper, we study in detail the prospects for detecting these objects, and show that GLAST can find a large fraction of the mini-spike population in the Milky Way over a time scale as short as 2 months, or set strong constraints on their existence in case of null searches. Finally we show that, in most cases, one can not only expect a clear discrimination from ordinary astrophysical sources, but also obtain a satisfactory determination of the DM mass as well as hints on its nature, through the determination of the leading annihilation channel. If the mass of the annihilating particle exceeds the energy range of GLAST, the study of the detected sources can be extended to higher energies by air Cerenkov telescopes (ACTs) like CANGAROO~\cite{canga}, HESS~\cite{hess}, MAGIC~\cite{magic} or VERITAS~\cite{veritas}.

This article is organized as follows. In Sec.~\ref{ind_det}, we give an overview over indirect DM searches through gamma rays and  introduce the gamma-ray spectra that are typically expected from the annihilation of DM particles. The main characteristics of GLAST, along with a brief account of its scientific goals and expected performances, are presented in Sec.~\ref{glast}. In Sec.~\ref{sensib}, we show the sensitivity maps for the detection and identification of point sources of DM annihilation. We then apply our results to mini-spikes around IMBHs in Sec.\ref{mini}, where we make detailed predictions about the detectability of these objects. Finally, we present our conclusions in Sec.~\ref{conclu}.

\section{Indirect dark matter searches through gamma rays}
\label{ind_det}

A theoretically particularly well-motivated type of DM candidates are weakly interacting massive particles (WIMPs) that appear in various extensions to the standard model of particle physics (SM); with masses and couplings at the electroweak scale, they would be thermally produced in the early universe and automatically acquire the necessary relic density to account for the DM today. Usually, the WIMP appears as the lightest of a whole set of new, heavy particles and its decay into SM degrees of freedom is protected by a new symmetry. 

The prototype example for a DM candidate of this type is the neutralino (see \cite{lsp} for a classic review) that appears in most supersymmetric extensions to the SM as the lightest supersymmetric particle (LSP) and is given by   
 a linear combination of the superpartners of the gauge and Higgs fields,
\begin{equation}
  \label{neut}
  \chi\equiv\tilde\chi^0_1= N_{11}\tilde B+N_{12}\tilde W^3 +N_{13}\tilde H_1^0+N_{14}\tilde H_2^0\,.
\end{equation}
The neutralino mass is usually several hundred GeV or less; for very high Higgsino or Wino fractions, however, it can be considerably higher (up to 2.2 TeV in the latter case).
More recently, considerable attention has also turned to models with universal extra dimensions (UED) \cite{Appelquist:2000nn}, where the higher-dimensional degrees of freedom manifest themselves as towers of new, heavy states in the low-energy, effectively four-dimensional theory. The lightest of these Kaluza-Klein particles (LKP) has a mass very close to the inverse of the compactification scale $R$, and for $R^{-1}\sim1\,$TeV it turns out to be a perfect WIMP DM candidate just as the neutralino \cite{Cheng:2002iz,LKP}. In the minimal UED setup \cite{Cheng:2002iz}, the LKP is given by the $\gamma^{(1)}$, the first KK excitation of the photon. For the discussion that follows, we will mainly have these DM candidates in mind; however, we will also argue that the gamma-ray signals we expect in these cases are rather generic, so that our conclusions about the prospects for a detection of DM annihilation point-sources will remain essentially unchanged even for other DM candidates.

Both LSP and LKP  are charged under a $Z_2$ symmetry ($R$-parity and KK parity, respectively), under which all SM particles remain neutral. This ensures their stability against decay, but at the same time allows the (pair-) annihilation into SM particles. The resulting differential  gamma-ray flux at earth is then given by
\be
  \label{flux}
	\Phi_{\gamma} (E) = \frac{\langle\sigma v\rangle}{8\pi m_\chi^3} \sum_fB^f\frac{\mathrm{d}N^f_\gamma}{\mathrm{d}x}\times\int_\mathrm{l.o.s} \mathrm{d}\ell(\psi) \rho^2(\mathbf{r})\,, 
\ee
where the integral over the DM density $\rho$ is along the line of sight for a given angle $\psi$ of observation, $\langle\sigma v\rangle$ is the total annihilation rate, $m_\chi$ the dark matter particle's mass and $E=m_\chi x$ the photon energy; the sum, finally, runs over all possible annihilation channels, with branching ratios $B^f$  and photon multiplicities $\mathrm{d}N^f_\gamma/\mathrm{d}x$.
The left-hand side of the above expression for $\Phi_{\gamma} (E)$ determines the \emph{spectral form} of the signal; it depends only on the underlying model of particle physics and can be computed to a very good accuracy. However, since the dark matter distribution $\rho(\mathbf{r})$ is only poorly known, the same is unfortunately not true for the line-of-sight integral that appears on the right-hand side of Eq.~(\ref{flux}). For classical targets of indirect DM searches like the galactic center, for example, these astrophysical uncertainties result in a predicted \emph{amplitude} for the flux that may vary over several orders of magnitude \cite{Fornengo:2004kj}. Another problem that indirect DM searches have to face is that regions of expected high DM density usually also contain a sizeable amount of ordinary matter which gives rise to more conventional, astrophysical sources of gamma-rays. This is particularly true for the galactic center and even if the
observed gamma-ray signal from that direction is still not identified
unambiguously, it is very unlikely that the HESS source  can mainly be
attributed to DM annihilations~\cite{Aharonian:2006wh}. 
A DM component could still be singled out wth GLAST in the 10 -
300 GeV energy range, where a DM source can provide a good fit of the 
EGRET data~\cite{GCcenter1,GCcenter2},
 but in  order to be conclusive, it is of great importance to provide
smoking-gun evidence for any claimed indirect detection of DM. Such
evidence may come from a multi-wavelength analysis of the annihilation
spectrum \cite{multiwave} or from pronounced spectral features that
cannot be mimicked by astrophysical processes, like line-signals
\cite{lines} or the very sharp cutoffs that are  associated with final
state radiation
\cite{Beacom:2004pe,Bergstrom:2004cy,Birkedal:2005ep,Bergstrom:2005ss}. A third possibility is the one that we will discuss later, namely the appearance of a great number of sources with identical spectra as expected in the mini-spike IMBH scenario.

\begin{figure}[t]
	\centering
       \psfrag{0.001}[][][0.8]{$0.001$}
       \psfrag{0.01}[][][0.8]{$0.01$}
       \psfrag{0.1}[][][0.8]{$0.1$}
       \psfrag{1}[][][0.8]{$1$}
       \psfrag{10}[][][0.8]{$10$}
       \psfrag{100}[][][0.8]{$100$}
       \psfrag{1000}[][][0.8]{$1000$}
       \psfrag{0.02}[][][0.8]{$0.02$}
       \psfrag{0.05}[][][0.8]{$0.05$}
       \psfrag{0.2}[][][0.8]{$0.2$}
       \psfrag{0.5}[][][0.8]{$0.5$}
       \psfrag{x}[t][][1.1]{$x=E/m_\chi$}
       \psfrag{y}[b][][1.1]{$\mathrm{d}N_\gamma/\mathrm{d}x$}
     \includegraphics[width=0.45\textwidth]{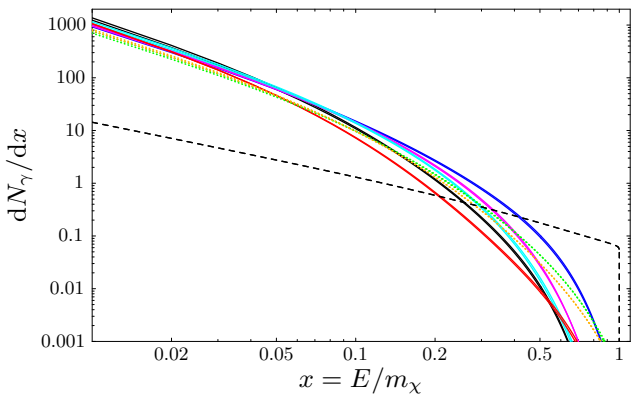}
     \caption{Photon multiplicity for various final states. The (blue,magenta,cyan,black,red) solid lines show the ($u\bar u$/$d\bar d$, $s\bar s$, $c\bar c$, $b\bar b$, $t\bar t$) quark spectra, while the (green,blue) dotted lines give the contributions from ($WW$, $ZZ$) gauge bosons. The (black) dashed line, finally, is the FSR spectrum from $e^+e^-$ final states. All spectra are plotted for both $m_\chi=500\,$GeV and $m_\chi=1000\,$GeV.}
     \label{fig_spectra1}
\end{figure}

Let us now investigate in a bit more detail the gamma-ray spectra that may result from DM annihilations. In principle, there are three different types of contributions to consider. The dominant source of gamma rays is usually associated to secondary photons from quark (or gauge boson) fragmentation and decay, mainly through the process $\pi^0\rightarrow\gamma\gamma$. The second contribution arises from final state radiation (FSR), where an additional photon is emitted from charged particle final states; this becomes particularly important for a sizable branching ratio into $e^+e^-$ pairs \cite{Beacom:2004pe,Bergstrom:2004cy}, though, for $m_\chi\gtrsim1\,$TeV, the same can be true even for $W^+W^-$ pairs \cite{Bergstrom:2005ss}. Finally, dark matter particles can annihilate directly into photons, giving rise to a characteristic line signal at the DM particle's mass \cite{lines}. However, since the DM has to be electrically neutral, these channels ($\chi\chi\rightarrow\gamma\gamma$ or $\chi\chi\rightarrow Z\gamma$) are necessarily loop-suppressed and thus usually negligible -- at least for current detector resolutions and in the absence of efficient enhancement mechanisms \cite{his}. Having in mind the limited reach of GLAST for very high energies, we therefore concentrate in the following on secondary photons and FSR photons from $e^+e^-$ final states. 

For the secondary photons, we have adopted the parameterizations that were obtained in Ref.~\cite{Fornengo:2004kj} by running the \textsc{Pythia} Monte Carlo code \cite{sjo}:
\begin{equation}
  \label{eq:param}
  \frac{\mathrm{d}N^f_\gamma}{\mathrm{d}x} = \eta x^a e^{b+cx+dx^2+ex^3} \,,
\end{equation}
where $\eta =2$ for the annihilation into WW, ZZ and $t \bar t$, and 1 otherwise.
The value of the parameters $(a,b,c,d,e)$ for the various annihilation channels $f$ and a set of DM masses can be found in Ref.~\cite{Fornengo:2004kj}.
We limit ourselves here to show in Figure~\ref{fig_spectra1} the photon 
multiplicities $\mathrm{d}N_\gamma/\mathrm{d}x$ per annihilation into 
quark and gauge boson pairs, for two different values of 
the DM mass, $M=500\,$GeV and $M=1000\,$GeV. For the $b \bar{b}$ channel, which we will  consider below, the parameter values are
$(\eta,a,b,c,d,e)=(1,1.50,0.37,-16.05,18.01,-19.50)$.

The $e^+e^-$ FSR spectrum, on the other hand, can be determined analytically and is, in leading logarithmic order, given by \cite{Bergstrom:2004cy,Birkedal:2005ep}
\begin{equation}
  \label{FSRf}
  \frac{\mathrm{d}N_\gamma^e}{\mathrm{d}x} \equiv
  \frac{\mathrm{d}(\sigma_{e^+e^-\gamma})/\mathrm{d}x}{\sigma_{e^+e^-}}
  \simeq \frac{\alpha}{\pi} \frac{(x^2 - 2x + 2)}{x} \ln{\left[
  \frac{m_\chi^2}{m^2_e}(1-x) \right]}\,.
\end{equation}
For the other charged leptons, one has simply  to replace the electron mass $m_e$ in the above expression with the corresponding lepton mass, leading to roughly the same spectral form as in the $e^+e^-$ case. The situation for $\tau$ leptons is actually a bit more complicated since their semi-hadronic decay produces secondary photons on top of the FSR contribution. The resulting spectrum exhibits a power-law behaviour for small photon energies, with a spectral index of about $-1.3$, and a cutoff that is sharper than in the case of quarks or gauge bosons, but less pronounced than for $e^+e^-$ pairs \cite{Fornengo:2004kj}. In the following, we will not consider this type of spectrum separately. We note that even in the LKP case, where gamma-rays from $\tau$ leptons are known to give an unusually important contribution \cite{Bergstrom:2004cy}, this does not change our general conclusions about the determination of the cutoff.

\begin{figure}[t]
	\centering
       \psfrag{0.001}[][][0.8]{$0.001$}
       \psfrag{0.01}[][][0.8]{$0.01$}
       \psfrag{0.1}[][][0.8]{$0.1$}
       \psfrag{1}[][][0.8]{$1$}
       \psfrag{10}[][][0.8]{$10$}
       \psfrag{100}[][][0.8]{$100$}
       \psfrag{1000}[][][0.8]{$1000$}
       \psfrag{0.02}[][][0.8]{$0.02$}
       \psfrag{0.05}[][][0.8]{$0.05$}
       \psfrag{0.2}[][][0.8]{$0.2$}
       \psfrag{0.5}[][][0.8]{$0.5$}
       \psfrag{x}[t][][1.1]{$x=E/m_\chi$}
       \psfrag{y}[b][][1.1]{$\mathrm{d}N^\mathrm{res}_\gamma/\mathrm{d}x$}
     \includegraphics[width=0.45\textwidth]{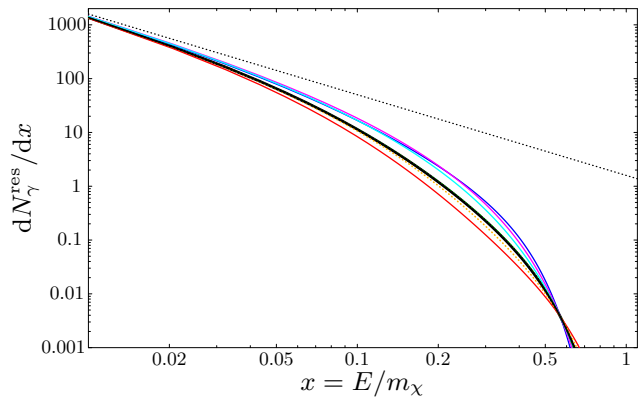}
     \caption{Same as Fig.~\ref{fig_spectra1} (for $m_\chi=500\,$ GeV and without the $e^+e^-$ FSR case), but with spectra rescaled as in (\ref{scaling}) and scaling parameters as given in Tab.~\ref{tab_scaling}.
The straight dotted line represents the extrapolated, asymptotic $x^{-1.5}$ behaviour of the photon spectrum from $b\bar b$ final states. The thick black line shows the (not rescaled) $b\bar b$ spectrum that we will use for reference.}
     \label{fig_spectra2}
\end{figure}

\begin{table}[b]
	\centering
   \begin{tabular}{|l||c||c|c|c|c|c|c|c|}
        \hline
           & $b\bar b$ & $u\bar u$ & $d\bar d$ & $s\bar s$ & $c\bar c$ & $t\bar t$ & $WW$ & $ZZ$ \\
        \hline \hline
        $A_f$ & 1.00 & 2.74 & 2.74 & 1.83 & 1.38 & 1.36 & 3.56 & 3.02 \\
        $B_f$ & 1.00 & 1.47 & 1.47 & 1.16 & 1.07 & 1.05 & 1.52 & 1.45  \\
        \hline
   \end{tabular}
     \caption{With these scaling parameters, the photon multiplicities for the  above final states are rescaled according to (\ref{scaling}), so as to feature the same asymptotic behaviour and cutoff as $b\bar b$ final states (here, we have defined the cutoff as a $10^{-3}$ drop of the spectrum with respect to the asymptotic $x^{-1.5}$ power law).\label{tab_scaling}}
\end{table}

We are now in a position to make two important observations about the gamma-ray spectra that can be expected from DM annihilations. First, expressed in the variable $x=E/m_\chi$, they are to a very good approximation independent of the DM mass $m_\chi$. Second, there are essentially only two types of spectra. FSR spectra, to begin with, show an asymptotic $x^{-1}$ power law for small $x$ and a characteristic, extremely sharp cutoff at the highest energy that is kinematically available. The spectra of secondary photons, on the other hand, exhibit an asymptotic $x^{-1.5}$ power law and a much less pronounced cutoff. In fact, spectra of this second type are almost indistinguishable if the dark matter mass is not known by some independent measurement. To make this point more clear, let us consider a rescaled version of the photon multiplicities for secondary photons,
\be
  \label{scaling}
  \frac{\mathrm{d}N^{f,\,\mathrm{res}}_\gamma}{\mathrm{d}x}(x)\equiv A_f\frac{\mathrm{d}N^\mathrm{f}_\gamma}{\mathrm{d}x}(B_f x)\,.
\ee
 \begin{figure*}[t]
  \centering
  \includegraphics[width=0.55\textwidth]{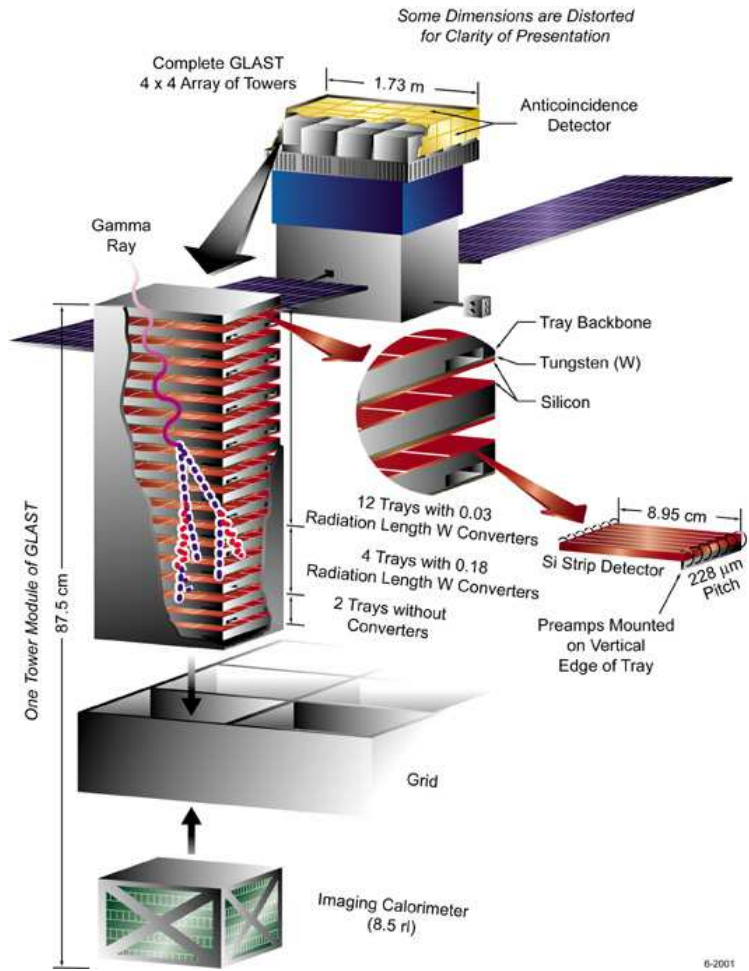}
  \caption{Schematic view of GLAST.}
  \label{fig:glast}
\end{figure*}
Here, $A_f$ represents an overall normalization of the amplitude while $B_f$ corresponds to a shift in the cutoff, i.e.~the DM mass. Since neither of these parameters is a priori known, we should use $\mathrm{d}N^{f,\,\mathrm{res}}_\gamma/\mathrm{d}x$, rather than $\mathrm{d}N^{f}_\gamma/\mathrm{d}x$, to compare the different spectra. The scaling parameters $A_f$ and $B_f$ can then be fixed by demanding that all spectra have the same normalization for small $x$ and the same cutoff (which we define here, somewhat arbitrary, as a $10^{-3}$ drop of the spectrum with respect to the asymptotic $x^{-1.5}$ power law). The resulting rescaled spectra are shown in Fig.~\ref{fig_spectra2}, with scaling parameters as given in Table~\ref{tab_scaling}.

{\it For illustration, we will in the following consider two different, rather simplified DM models: in model $A$, the DM mainly annihilates into $b\bar b$ pairs while in model $B$, we fix a branching ratio of 80\% into $b \bar b$ and 20\% into $e^+e^-$ final states}. The first case corresponds roughly to the situation one generally expects for the LSP (at least for masses smaller than a few hundred GeV, see, e.g., \cite{Bertin:2002ky}), while a large branching ratio into leptons like in model $B$ is for example realized in the case of the LKP. From the above discussion, however, it should be clear that our models $A$ and $B$ are more general in the sense that they cover the two basic possibilities for spectral distributions that can be expected from DM annihilations.
 At this point, a word of caution is in order. While the characteristic sharp FSR cutoff can be used to unambiguously identify the DM mass to an accuracy that corresponds to the energy resolution of the detector, a certain theoretical uncertainty enters into the determination of $m_\chi$ when only the rather soft cutoff from secondary photons is observed. If, e.g., DM mainly annihilates into light quarks or gauge bosons, one would not be able to tell the difference from our case $A$, but  underestimate the actual DM mass by up to 50\% (see Table~\ref{tab_scaling}).

Let us conclude this Section by stressing once more that the detection of many sources with the same spectrum, and in particular with the \emph{same cutoff}, would provide a smoking-gun evidence for DM annihilations -- even if the cutoff is not very pronounced, like in model $A$, and would therefore not provide conclusive evidence when observed from a single source alone.

\section{GLAST} 
\label{glast}

The Gamma-ray Large Area Space Telescope (GLAST) is a next generation gamma ray observatory due for launch in Fall 2007. The main scientific objects are the study of all gamma ray sources such as blazars, gamma-ray bursts,  supernova remnants, pulsars, diffuse radiation, and unidentified high-energy sources. Indeed, GLAST might be called the "Hubble Telescope" of gamma-ray astronomy as it will be able to observe AGN sources up to z~$\sim$~4  and beyond, if such objects actually existed at such early times in the universe.   Extrapolation from EGRET AGN detections shows that about 5,000 AGN sources will be detected in a 2 years cumulative scanning mode observation by GLAST, as compared to the 85  that have been observed by EGRET in a similar time interval. The GLAST collaboration has also a strong interest in DM detection and several works in the literature investigate possible scenarios and the corresponding detection capabilities (see e.g. Ref.~\cite{DMlat} and the aforementioned reviews ~\cite{review1,review2} for further details).

The primary instrument on board is the Large Area Telescope (LAT), a pair-production telescope designed to allow observation of celestial sources in the range 20~MeV--300~GeV \cite{LAT}; LAT was developed by an international collaboration of  particle physics and astrophysics communities from 26 institutions in the United States, Italy, Japan, France and Germany. The LAT is an  4x4 array of identical towers, each formed by
\begin{itemize}
\item   Si-strip Tracker Detectors and converters arranged in 18 XY tracking planes for the measurement
of the photon direction,
\item Segmented array of CsI(Tl) crystals for the measurement the photon energy,
\item Segmented Anticoincidence  Detector (ACD).
\end{itemize}
\begin{table}[t]
\centering
  \begin{tabular}{|l|c|}
    \hline
    Peak Effective Area (1--10 GeV)&$>$8000~cm$^2$\\
    Energy Resolution 10 GeV on-axis&$<$10\%\\
    Energy Resolution 10-300 GeV on-axis&$<$20\%\\
    Energy Resolution 10-300 GeV $>$60$^\circ$&$<$6\%\\
    PSF 68\% 10~GeV on-axis&$<0.15^\circ$\\
    Field of view&2.4~sr\\
    Point source sensitivity ($>$100MeV,5y)&$<6 \times 10^{-9}$~cm$^{-2}$s$^{-1}$\\
    Source Location&$<0.5$~arcmin \\
    \hline	
  \end{tabular}
  \caption{GLAST LAT minimum performances, adapted from Ref.~\cite{LATspec}.\label{tab:lat}
}
\end{table}

During normal operation the LAT will continuously scan the whole sky, obtaining a complete coverage about every 3 hours due to the great field of view ($\sim 2$~sr). The uniform sky coverage, in addition to the large effective area and the good angular resolution should ensure many advances in the study of all celestial sources of interest. In Tab.~\ref{tab:lat}, we summarize from \cite{LATspec} the LAT performances more relevant in our context.

\section{Dark Matter Sensitivity Maps}
\label{sensib}

In order to study the LAT sensitivity for DM annihilation signals, we perform a simulation of the gamma-ray sky as it will be observed by this instrument, based on a parametrization of the instrument response. This allows us to rapidly simulate several months of data-taking in a relatively short time, without having to deal with a complex MonteCarlo simulation of the physical processes involved. For additional information about GLAST LAT simulations, see, for example, Ref.~\cite{Gsim}.

All simulations performed for this study correspond to 2 months of scanning (i.e.~not pointed) LAT observations. The time duration was selected such as to allow for a relatively fast discovery of a DM signal; the trade-off is a sizeable decrease in the number of resolved and correctly identified sources. No other (non-DM) {\it point} sources were simulated and source spatial coincidence was not considered, in order to simplify the analysis and to allow for an automated analysis process. 

The extragalactic diffuse background was simulated by extending the EGRET observations to the LAT energy range~\cite{EGdiff}. The galactic diffuse background model is implemented in the framework of the GALPROP code for cosmic-ray propagation and incorporates up-to-date surveys of the interstellar medium, as well as current models for the interstellar radiation field, updated production functions and inverse scattering calculations~\cite{LATdiff}.
As motivated in Sec.~II, we will focus on DM candidates mainly annihilating into $b \bar{b}$, and discuss below how the detection of deviations from the corresponding gamma-ray spectrum may provide useful insights on the nature of the DM particles. 

\begin{figure*}[t]
   \centering
   \includegraphics[scale=0.8,angle=270]{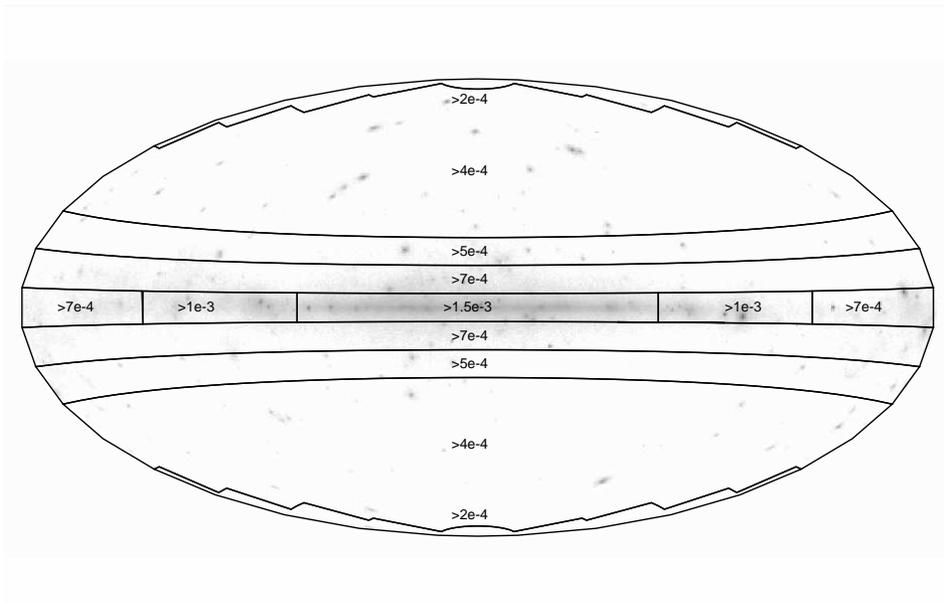}
   \caption{GLAST sensitivity map for the {\it detection} of point sources of Dark Matter annihilations, i.e. full-sky map in galactic coordinates of the minimum flux above 20 MeV, in units of [ph m$^{-2}$s$^{-1}$], that is required for a $5\sigma$ detection of an annihilation spectrum, assuming a DM particle with mass $m_\chi=150~$GeV annihilating into $b \bar b$ (note, however, that the map does not depend very sensitively on DM properties). The map is relative to a 2 months observation period; for longer observation times, fluxes scale approximately as $t_{obs}^{-1/2}$. For reference, we also show the simulated gamma ray sky. }
   \label{sensmap}
\end{figure*}
\begin{figure*}[t]
   \centering
   \includegraphics[scale=0.8,angle=270]{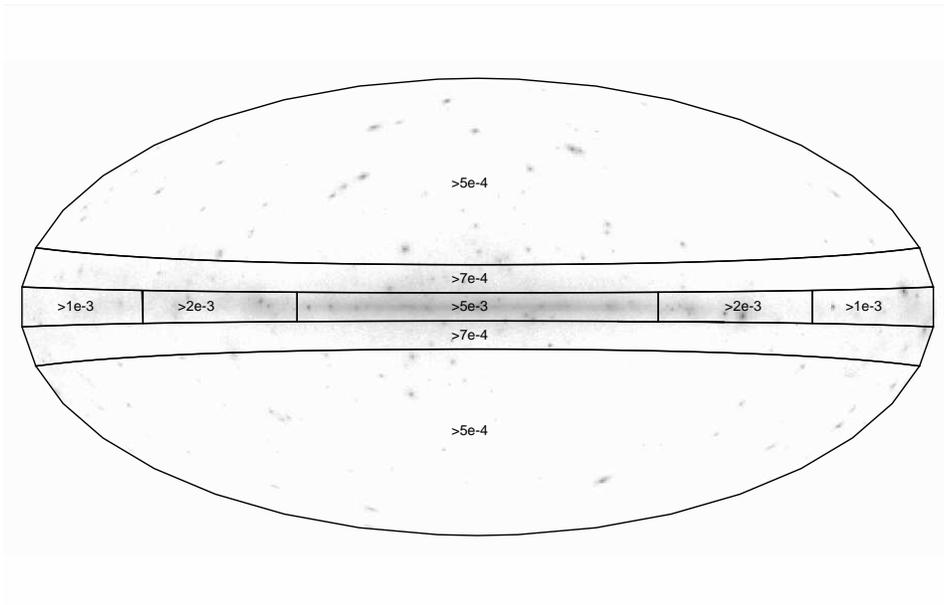}
   \caption{GLAST sensitivity map for the {\it identification} of point sources of Dark Matter annihilation, i.e. full-sky map in galactic coordinates of the minimum flux above 20 MeV, in units of [ph m$^{-2}$s$^{-1}$], necessary to discriminate the annihilation spectrum from an ordinary astrophysical source, assuming a DM particle with mass $m_\chi=150~$GeV annihilating into $b \bar b$, after a 2 months observation period (see text for a further discussion). For reference, we also show the simulated gamma ray sky.}
   \label{cutmap}
\end{figure*}

In the first part of our analysis, we investigated the intensity required to resolve a point-source of DM annihilations at a $5\sigma$ significance level. We divided the sky into regions of about 10 degrees in radius, and in each region we placed one DM source. Then, we considered each source separately and let the flux intensity vary from $10^{-4}$ to $2\times 10^{-3}$ ph m$^{-2}$s$^{-1}$. For each intensity, we calculated the significance of the observed signal, given the local background counts, with a maximum likelihood analysis assuming Poisson statistics. By estimating the minimum flux required to discriminate the DM source from the background at a $5\sigma$ level on a grid of points uniformly distributed over the sky, we have obtained the sensitivity map shown in the top panel of Fig.~\ref{sensmap} (where we adopted a DM particle mass $m_\chi = 150$~GeV). The sensitivity appears to depend significantly on the Galactic longitude only along the Galactic disk, as expected. At high galactic latitudes a source as faint as $2\times 10^{-4}$ ph m$^{-2}$s$^{-1}$ is resolved, while close to the galactic center a minimum flux of $1.5\times 10^{-3}$ ph m$^{-2}$s$^{-1}$ is required.

In a similar way, we can also produce a sensitivity map for a reliable identification of the DM spectral cutoff. To do so, we compare for each observed spectrum the likelihood of a plain power law, with that of a DM-like spectrum, and check whether the latter hypothesis is favoured. The result of this analysis is shown in Fig.~\ref{cutmap}. Inside the galactic plane ($\vert b\vert<5$) the minimum intensity required to pinpoint the high energy cutoff is far greater than the intensity required for a $5\sigma$ detection. This implies that there will be some sources that can correctly be identified as a DM signal only after observation times longer than the 2 months considered here.

All figures quoted in this work were calculated by taking into account in the analysis only detected photons with energies between 100 MeV and 200 GeV. We limit ourselves to this energy range in order to avoid the (present) uncertainties in the LAT response functions at very low (down to $\sim 20\,$MeV) and high energies (up to $500\,$GeV). On the other hand, we do take into account all major effects such as orbiting, rocking and dead times.
While our sensitivity maps are in principle still subject to further revisions as the knowledge of the LAT response improves, we are  therefore confident that they will receive only  minor corrections (\emph{improving} the detection efficiency).

The time scale of 2 months adopted here corresponds roughly to the periodicity of the LAT orbit given an inclination of $\sim28.5^\circ$ ($\sim55$ days). Assuming that the detection statistics are dominated by diffuse counts and not by stochastic fluctuation, it is possible to build {\it detection} maps for longer observation times $t_{obs}$ by simply scaling down the minimum flux with $t_{obs}^{-1/2}$. For instance, by increasing the observation time by a factor $4$, $t_{obs}=8$ months, the minimum fluxes shown in the maps are roughly halved. For the identification maps, on the other hand, the situation is more complicated. It is in fact not trivial to scale minimum fluxes for the determination of the DM mass to longer observation time, since the analysis is based on a comparison of likelihood between models, in a regime where both background counts and low statistics play an important role. Although a scaling $t_{obs}^{-1/2}$ appears as a conservative estimate in this case, we feel that this issue deserves a dedicated study.

The sensitivity map for the {\it detection} of DM point sources is rather insensitive to the DM particle mass, since the flux is dominated by the low energy tail of the annihilation spectrum. However, things are different for the {\it identification} map, at least above $\sim 200$GeV. In fact, the determination of a high energy cut-off strongly depends on the possibility of obtaining good quality data (in terms of angular and energy resolution), for photons above this energy. A detailed study of high-mass models will thus be possible only with a reliable simulation of the instrument response at energies above the range considered, although it should be possible to identify its DM origin, through the rather distinctive slope of the annihilation spectrum (see discussion below). Finally, we note that neither the detection nor the identification of DM point sources depends significantly on the annihilation channels; this situation would, in principle, change for longer observation times and DM masses below a few hundred GeV, when the pronounced sharp cutoff for FSR photons from $e^+e^-$ final states becomes clearly visible, thereby enhancing the prospects for the identification of model $B$ type DM.

We stress that
{\it the sensitivity maps in Fig.~\ref{sensmap} and \ref{cutmap} represent a powerful tool 
to assess the observability of} any {\it population of point sources of DM annihilations,
without having to perform a case-by-case study (simulation plus analysis) of individual 
sources}.

\section{Detectability of Mini-Spikes}
\label{mini}

We now apply the results obtained in the first part to a specific
astrophysical scenario, namely to DM mini-spikes around Intermediate Mass Black Holes (IMBHs), and study the prospects
for detecting these objects with GLAST. In this section, following closely
 Ref.~\cite{Bertone:2005xz}, we will first 
review the basics of the mini-spike
scenario, and derive, for a Milky-Way like galaxy, their number, 
profile and total luminosity in terms of DM annihilations. 
We will then combine this information with the sensitivity maps shown above to 
determine the prospects for a detection with GLAST.

\subsection{The Mini-spikes scenario}
The effect of the formation of a central 
object on the surrounding distribution of matter has been investigated 
in Refs.~\cite{peebles:1972,young:1980,Ipser:1987ru,Quinlan:1995} 
and for the first time in the framework of DM annihilations in 
Ref.~\cite{Gondolo:1999ef}. It was shown that  the 
{\it adiabatic} growth of a massive object at the center of a 
power-law distribution of DM with index $\gamma$, induces 
a redistribution of matter into a new power-law (dubbed ``spike'') with index
\begin{equation}
\gamma_{sp}=\frac{9-2\gamma}{4-\gamma}\,.
\end{equation}
This formula is valid over a region of size $r_{sp} \approx 0.2 r_{h}$, 
where $r_{BH}$ is the radius of gravitational influence
of the black hole (see below for further details).
The adiabaticity of the black hole's growth is in particular valid for the SMBH at the 
Galactic center. A critical assessment of the formation {\it and survival}
of the central spike, over cosmological timescales, is presented
in Refs.~\cite{Bertone:2005hw,Bertone:2005xv} (see also references therein).
We limit ourselves here to note that adiabatic spikes are rather
fragile structures, that require fine-tuned conditions to form 
at the center of galactic halos~\cite{Ullio:2001fb},
and that can be easily destroyed by dynamical processes
such as major mergers~\cite{Merritt:2002vj}
and gravitational scattering off stars~\cite{Merritt:2003eu,Bertone:2005hw}. 

It was recently shown that a $\rho \propto r^{-3/2}$ DM overdensity can
be predicted in any halo at the center of any galaxy old enough to 
have grown a power-law density cusp {\it in the stars} via the 
Bahcall-Wolf mechanism~\cite{Merritt:2006mt}. Collisional generation of these 
DM "crests" (Collisionally REgenerated STtructures) was demonstrated 
even in the extreme case where the DM density was lowered by 
slingshot ejection from a binary super-massive black hole. However, 
the enhancement of the annihilation signal from a DM crest is 
typically much smaller than for adiabatic spikes~\cite{Merritt:2006mt}.

Although it is unlikely that a spike may survive around the 
Super-massive Black Hole at the Galactic center, they can evolve 
unperturbed around Intermediate Mass Black Holes (IMBHs), i.e.
wandering BHs with mass $10^2 \lesssim M/M_\odot \lesssim 10^6$. 
Scenarios that seek to explain the properties of the observed 
super-massive black holes population  result, in fact, in the prediction
of a large population of IMBHs. The number and properties of 
mini-spikes around IMBHs have been discussed in 
Ref.~\cite{Bertone:2005xz}, where two different formation scenarios 
have been investigated. In the first one, IMBHs form in rare, 
overdense regions at high redshift, 
$z \sim 20$, as remnants of Population III stars, and have a 
characteristic mass-scale of a few $10^2 M_\odot$ 
\cite{Madau:2001} (the formation of mild mini-spikes around
these objects was also discussed in 
Ref.~\cite{Zhao:2005zr,islamc:2004,islamb:2004}).  
In this scenario, these black holes serve as the 
seeds for the growth of super-massive black 
holes found in galactic spheroids \cite{Ferrarese:2005}. We will not 
further consider this scenario, because of the many uncertainties 
associated to the initial and final mass function of IMBHs, and 
the lack of a solid prescription for the initial DM distribution
and BH mass grow due to gas accretion as these objects move in 
galactic halos. 

We will instead focus on the second scenario discussed in 
Ref.~\cite{Bertone:2005xz}, based on the proposal 
of Ref.~\cite{Koushiappas:2003zn}, which is representative of a class of models
in which black holes originate from 
massive objects formed directly during the 
collapse of primordial gas in early-forming 
halos. In practice, during the virialization and collapse
of the first halos, the gas cools and forms pressure-supported disks 
at the centers of halos that are sufficiently massive 
for molecular hydrogen cooling to be efficient. 
Gravitational instabilities in the disk 
lead to an effective viscosity that transfers 
mass inward and angular momentum outward 
until the first generation of stars heats the disk and terminates 
this process.  In this case, the 
characteristic mass of the black hole forming in a halo of virial mass 
$M_v$ is given by ~\cite{Koushiappas:2003zn}
\begin{eqnarray}
\label{eq:mbh}
M_{\rm bh}& = 3.8 \times 10^4 M_\odot 
\left( \frac{\kappa}{0.5} \right)
\left( \frac{f}{0.03} \right)^{3/2} \nonumber \\
& \left( \frac{M_v}{10^7 M_\odot} \right)
\left( \frac{1+z}{18} \right)^{3/2}
\left( \frac{t_{ev}}{10 {\rm Myr}} \right),
\end{eqnarray}
where $f$ is the fraction  
of the total baryonic mass in the halo that has fallen into 
the disk, $z$ is the redshift of formation, $\kappa$ 
is the baryon fraction that lost its angular momentum 
on a time $t_{ev}$, and $t_{ev}$ is the timescale for the evolution of the 
first generation of stars \cite{Koushiappas:2003zn}.  
The distribution of black hole masses is a log-normal 
distribution with a mean value given by the characteristic 
mass above and a standard deviation $\sigma_{M_{\rm bh}}=0.9$.  
  
The ``initial'' DM host mini-halo (that is, the DM distribution prior 
to black hole formation) can be well approximated by a 
Navarro, Frenk, and White (NFW) profile~\cite{Navarro:1996he}
\begin{equation}
\rho(r)=\rho_0 \left( \frac{r}{r_s} \right)^{-1} \left( 1+\frac{r}{r_s} \right)^{-2}\,.
\label{eq:nfw}
\end{equation}
The normalization constant $\rho_0$, and the scale radius $r_s$, can
be expressed in terms of the virial mass of the halo at the time when
the IMBH formed, $M_{{\rm vir}}$, and the virial concentration parameter, 
$c_{{\rm vir}}$. See Ref.~\cite{Bertone:2005xz} for further details.
Alternatively, we could have chosen the more 
recent parametrization proposed
by Navarro et al.~\cite{Navarro:04b} (see also
Refs.~\cite{Reed:05,Merritt:05}).  However, this profile 
implies modifications at scales smaller than those we are 
interested in, where the profile is 
anyway modified by the presence of the IMBH, hence leading to a negligible
change in the predicted fluxes.

We assume that the black holes form over
a timescale long enough to guarantee adiabaticity, 
but short compared to the cosmological evolution of 
the host halo (see Ref.~\cite{Bertone:2005xz} for 
further details).  In fact, 
the condition of ``adiabaticity'', fundamental to 
grow ``mini-spikes'', requires that the formation time of 
the black hole is much larger than the dynamical 
timescale at a distance $r_h$ from the black hole, 
where $r_h$ is the radius of the sphere of gravitational 
influence of the black hole, 
$r_h \simeq G M_{\rm bh}/\sigma^2$,
and $\sigma$ is the velocity dispersion of 
DM particles at $r_h$.
In practice, $r_h$ is estimated by solving 
the implicit equation 
\begin{equation}
  \label{mrh}
M(<r_h) \equiv \int_{0}^{r_h} \rho(r) r^2 \; {\rm d}r = 2 \; M_{\rm bh}\;\; .
\end{equation}
Inside a region $r_{sp}\approx 0.2 r_h$~\cite{Merritt:2003qc}, 
and assuming  a NFW profile
as in Eq.~(\ref{eq:nfw}), the spike profile is  given by
\begin{equation}
\rho_{sp}=\rho \left( r_{sp} \right) \left( \frac {r}{r_{sp}} \right)^{-7/3}\,.
 \end{equation}
\begin{figure*}[t]
   \centering
   \includegraphics[width=0.65\textwidth]{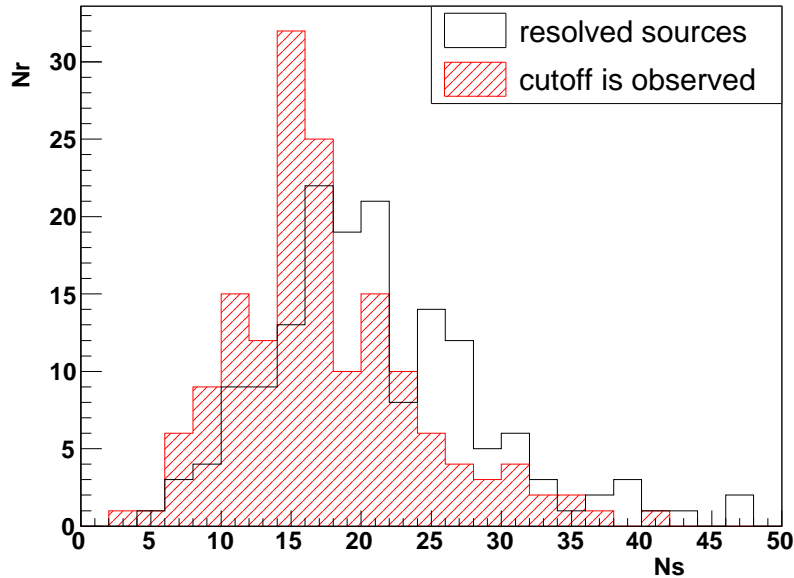}
   \caption{Number of realizations $N_R$ where a number $N_S$ of mini-spikes is observed (empty histogram) and identified as DM sources (red dashed), assuming $m_\chi=150$~GeV.}
   \label{real1hist}
\end{figure*}
\begin{table*}[t]
  \centering
  \begin{tabular}{|l|c|c|c|}
    \hline
    &Accepted realizations&Resolved mini-spikes&Identified mini-spikes\\\hline
    Standard scenario &159/200&$22\pm8$&$19\pm7$\\
	\hline
    Scaled $\xi=0.2$ &198/200&$4\pm2$&$4\pm2$\\
    Scaled $\xi=5$ &13/200&$52\pm14$&$45\pm12$\\
    \hline
  \end{tabular}
  \caption{Statistics of the number accepted realizations (i.e. those not violating the EGRET constrain), resolved sources (i.e. those detected by GLAST within 2 months at 5$\sigma$) and identified sources (i.e. those that can be discriminated by ordinary astrophysical sources), for M=150~GeV, and 2 different values of the rescaling factor $\xi$ that accounts for a variation of the astrophysical and particle physics parameters in the mini-spike model.\label{stats}}
\end{table*}
Such a steep power-law leads to a density profile that diverges at small radii. Although
the Schwarzschild radius of the black hole provides already a physical 
lower cut-off, the profile already 'saturates' at larger radii, where the DM 
annihilation timescale becomes shorter than the time $t_{sp}$ elapsed since 
the formation of the spike. This radius $r_{\rm lim}$ can be expressed implicitly 
as
\begin{equation}
\rho_{\rm sp}(r_{\rm  cut})=m_\chi/\sigma v \,(t-t_{sp}) 
\,,
\label{eq:rholim}
\end{equation}
where $\sigma v$ is the annihilation cross section times the relative velocity
of the DM particle. When performing our integrations, we thus stop at a ``cut'' radius 
defined as
\begin{equation}
r_{\rm cut}(m,\sigma v)={\rm Max} \left[ 4 R_s, r_{\rm lim}(m,\sigma v) \right]\,,
\end{equation}
where $R_s = 2.95 \,{\rm km} \, M_{\rm bh}/M_\odot$ is the Schwarzschild radius of the 
IMBH.  

The flux of gamma-rays from a mini-spike around an IMBH can be 
expressed as ~\cite{Bertone:2005xz}
\begin{eqnarray}
\nonumber
\Phi (E,D) & = & \Phi_0 \frac{{\rm d}N}{{\rm d}E} 
\left( \frac{\sigma v}{10^{-26} {\rm cm}^3/{\rm s}} \right)
\left( \frac{m_\chi}{100 {\rm GeV}} \right)^{-2} 
\\ 
 & \times & \left( \frac{D}{{\rm kpc}} \right)^{-2} 
\left( \frac{\rho(r_{\rm sp})}{10^2 {\rm GeV}{\rm cm}^{-3}} \right)^{2} 
\nonumber \\
 & \times & 
\left( \frac{r_{\rm sp}}{{\rm pc}} \right)^\frac{14}{3}
\left( \frac{r_{\rm cut}(m,\sigma v)}{10^{-3}{\rm pc}} \right)^{-\frac{5}{3}}\, ,
\label{eq:flux}
\end{eqnarray}
where $\Phi_0 = 9 \times 10^{-10} {\rm cm}^{-2}{\rm s}^{-1}$  
and $D$ is the IMBH distance to the Earth. Note that the annihilation 
flux in this case does not depend, as one may na\'ively expect, 
on the ratio $(\sigma v/m_\chi^2)$, because $r_{\rm cut}$ is itself a 
function of the particle mass and cross section. It is easy to show that in
this case the (integrated) gamma-ray flux is proportional to 
$\sim (\sigma v)^{2/7} m_\chi^{-9/7}$ (given that $dN/dx$ is almost independent of $m_\chi$), which means that -- due to the ``saturation'' of the DM profile described above -- the predicted
fluxes are much less sensitive to the particle physics parameters than in 
the usual case. 

In order to assess the detectability of mini-spikes, we have followed the procedure 
outlined in Refs.~\cite{Bertone:2005xz,Bertone:2006nq} to build mock catalogs of sources. As detailed in the next Section, we 
have then performed our analysis on 200 {\it statistical realizations} of a Milky-Way like 
halo, studied the detectability in each realization, and then performed a statistical
analysis of the results.

\begin{figure*}[t]
   \centering
	\subfigure{
   \includegraphics[width=0.4\textwidth]{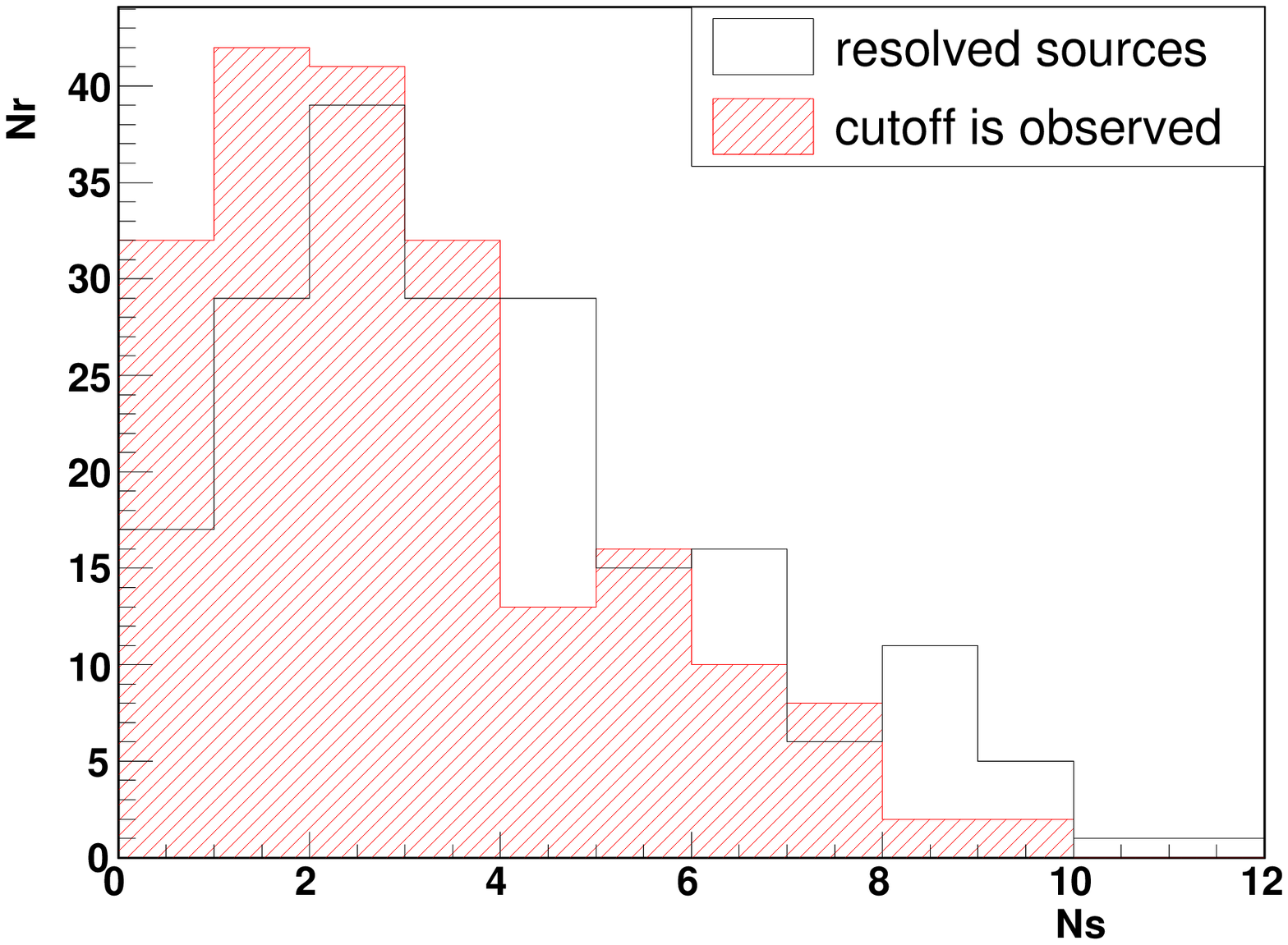}}
	\hspace{0.5cm}
	\subfigure{
   \includegraphics[width=0.4\textwidth]{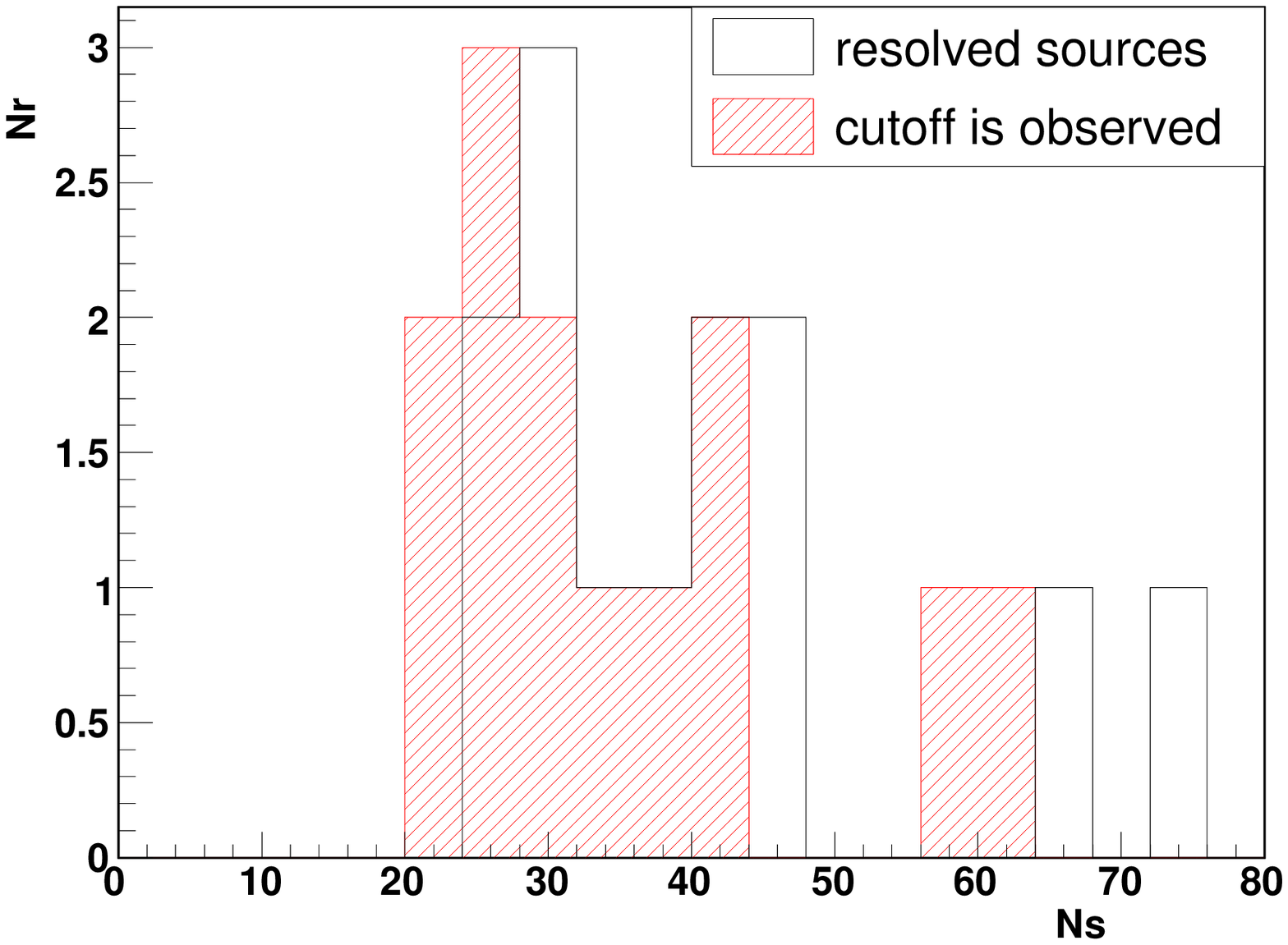}}
   \caption{Number of realizations $N_R$ where a number $N_S$ of mini-spikes is observed (empty histogram) and identified as DM sources (red dashed), assuming $m_\chi=150$~GeV, and a rescaling factor $\xi=0.2$ (left) and $\xi=5$ (right).}
   \label{reals}
\end{figure*}
\subsection{Prospects for detection with GLAST}

We now turn to the prospects for detecting mini-spikes with GLAST. To this aim, we simply have to compare our mock catalogs of sources, described in more detail in the previous section,  with the sensitivity maps of Figs.~\ref{sensmap} and \ref{cutmap}. If the flux from a given source is greater than the minimal flux required to grant a $5\sigma$ detection we will say that the source is {\it resolved}; if it is also greater than the threshold for the discrimination from astrophysical sources, we will say the source is {\it identified}.  Let us point out here that, obviously, the discussion on the detectability of mini-spikes can
only be performed in a statistical sense, since we do not know in what realization 
of the simulated Milky Way halo we live in.

The result of the above procedure is shown in Fig.~\ref{real1hist}, where we show the number $N_R$ of
realizations with  $N_S$ resolved (and identified) mini-spikes. Let us stress that some of the realizations actually contain sources that would shine brighter than the brightest EGRET 
unidentified source outside the galactic plane (3EG J1835p5918, \cite{3EGcat}).
In order not to violate the observational constraints, we therefore reject 
any realization where the flux of at least one source exceeds 
$\Phi_{\rm max}=2 \times 10^{-2}$  ph m$^{-2}$ s$^{-1}$ above $20$~MeV; these rejected realizations are \emph{not} included in Fig.~\ref{real1hist} and will not be considered in the following, either.
Incidentally, it is interesting to note that the spectral index reported for the brightest unidentified EGRET source is $n=1.69\pm 0.07$, quite close to what is expected for the low energy tail of the annihilation spectrum (EGRET was sensitive to energies up to 30~GeV). At lower energies, COMPTEL (in the range 0.75--30~MeV) did not resolve this source from the steep blazar QSO1739+522 {\cite{comptel}}. 

We can see from Fig.~\ref{real1hist}
that, in most realizations and adopting a particle mass $m_\chi=150$ GeV, we can expect a large number of sources 
 to be detectable; Tab.~\ref{stats}, furthermore, shows that only a small fraction of all the realizations 
is in violation of the observational constraints set by EGRET. 

The expected flux from any given mini-spike does, however, depend on various particle physics 
and astrophysical parameters. For what concerns the particle physics parameters, we
have already stressed that $\Phi_i(E) \propto (\sigma v)^{2/7} m_\chi^{-9/7}$.
As for the astrophysical uncertainties, we have shown in Eq.~\ref{eq:mbh}  the dependence of the 
BH mass on the formation model parameters. Without entering here into a detailed discussion
of the parameter space of the model, we incorporate the effect of varying all parameters,
by introducing a global rescaling parameter $\xi$, building new realizations with 
fluxes 
\begin{equation}
\Phi_{\xi}(E,D)=\xi \, \Phi_i(E,D) \,\,,
\end{equation}
where $\Phi_i$ are the fluxes obtained in the 
reference standard scenario discussed above, which then corresponds to the case $\xi=1$.
For high values of $\xi$, we expect a lower number of acceptable sources, since many 
of them will violate the EGRET constraint, while lowering $\xi$ will make a larger number 
of realizations compatible with the same bound. 
We show the results for the cases $\xi=0.2$ and $5$, respectively, in Fig.~\ref{reals}, and 
summarize the number of accepted realizations, 
resolved sources and identified sources, in Tab.~\ref{stats}. 
\begin{figure*}[t]
   \centering
	\subfigure[$\phi=2\times 10^{-3}$ ph m$^{-2}$ s$^{-1}$, $m_\chi =150$GeV, $b \bar b$, ($l,b$)=($0,25$)]{ 
   \includegraphics[width=0.3\textwidth]{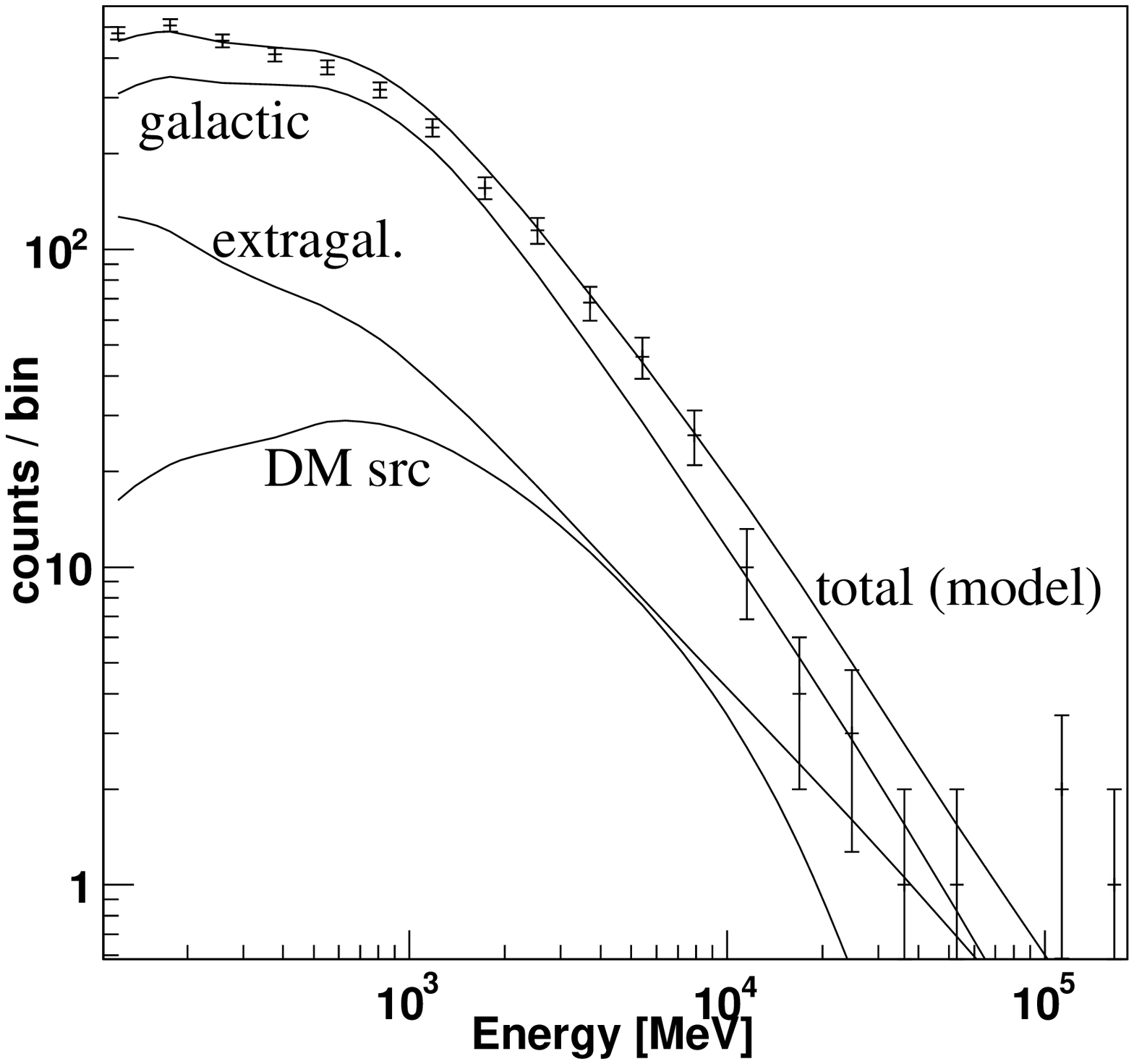}}
	\subfigure[$\phi=2\times 10^{-2}$ ph m$^{-2}$ s$^{-1}$, $m_\chi =150$GeV, $b \bar b$, ($l,b$)=($50,0$) ]{ 
   \includegraphics[width=0.3\textwidth]{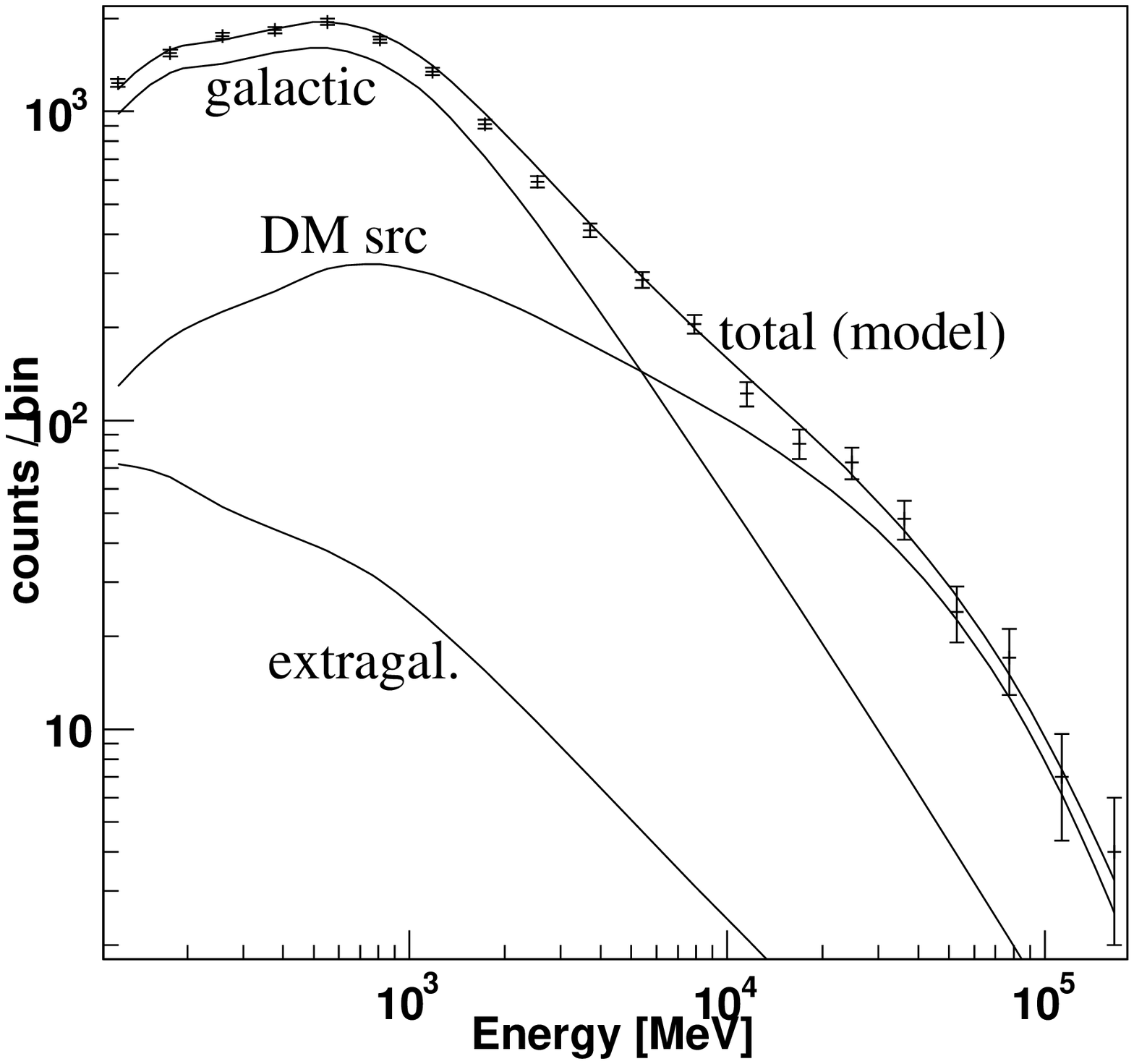}}
	\subfigure[$\phi=2\times 10^{-2}$ ph m$^{-2}$ s$^{-1}$, $m_\chi =150$GeV, 80 \% $b \bar b$, 20 \% $e^+ e^-$,  ($l,b$)=($0,50$)]{ 
   \includegraphics[width=0.3\textwidth]{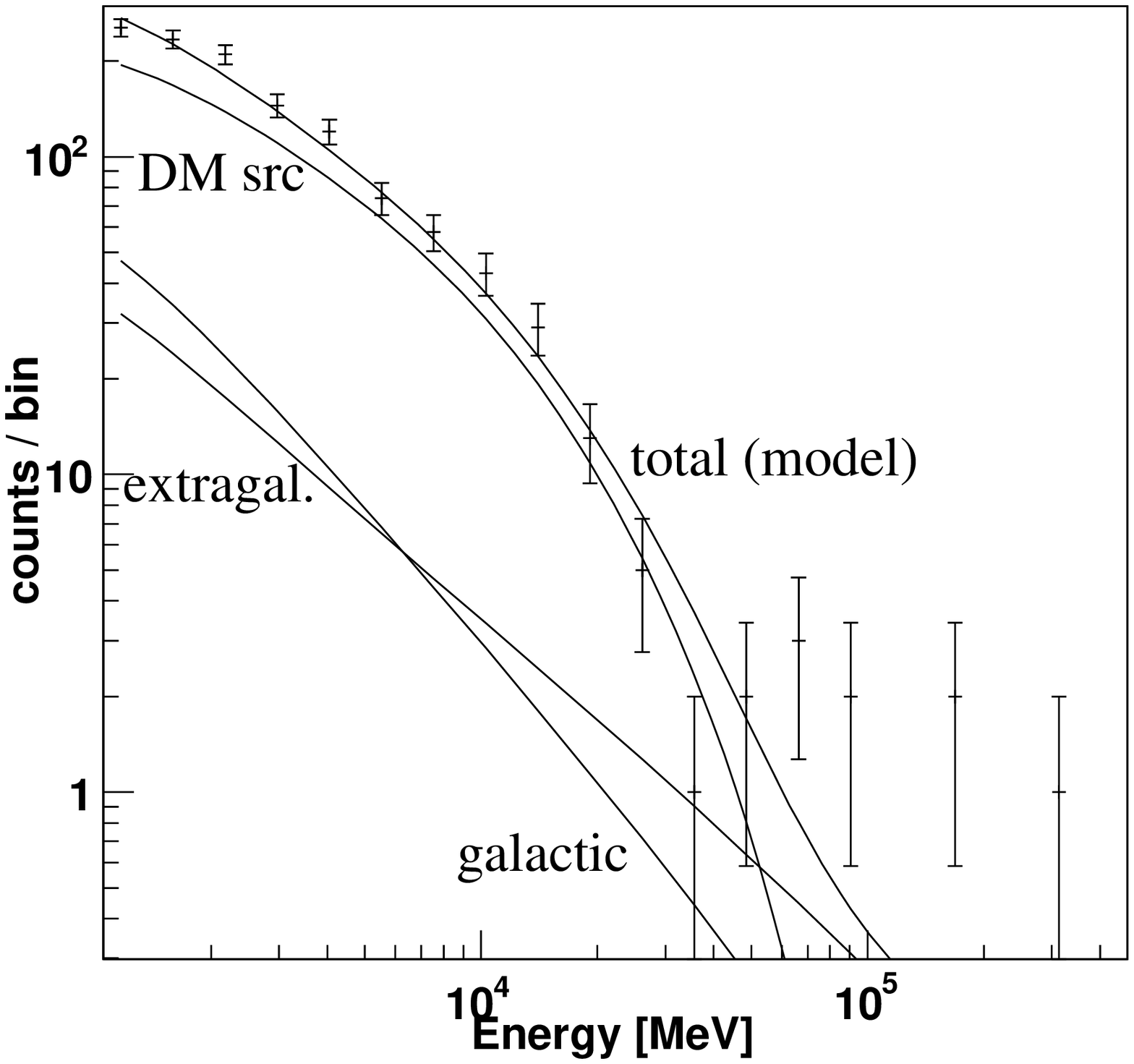}}
   \caption{Examples of spectral fits of simulated DM point sources of intensity $\phi$, for different values of $m_\chi$ and different annihilation channels. Solid lines are fits obtained under the assumption of annihilation to $b \bar b$. For each model we also give the significance of the detection. Points with error bars are photon counts from the simulated observation.}
   \label{spectra}
\end{figure*}

\section{Towards the nature of Dark Matter}
\label{sec:DMnature}

In this Section, we analyze in more detail some aspects of the detected sources.  As a first step, it is important to verify under which conditions the LAT is able to detect the cutoff for a DM candidate like, e.g., in our model $A$, i.e. under which conditions the fit obtained by adopting the parametrization of Eq.~(\ref{eq:param}) (with only $m_\chi$ and the absolute normalization as free parameters) has a greater likelihood than an ordinary power-law fit (where the free parameters are the overall normalization and the spectral slope).
We find that the highest DM mass $m_\chi$ that still gives a spectrum recognizable as DM-induced is close to 1 TeV for the brightest sources. The efficiency in the $m_\chi$ estimate, however, is strongly dependent on the mass itself. For a simulated source with $m_\chi=150\,$GeV, e.g., we can estimate the mass of the DM particle with a precision better than 25 GeV. For higher masses the precision on the determination of the DM mass depends strongly on the possibility to extend the instrument's energy range to photon energies above 200~GeV; an educated guess on the instrument's response suggests that a mass of 1~TeV can be reconstructed with a precision of $~10$\% for the brightest sources. A detailed analysis would warrant a dedicated study, also because the exact role of systematics and stochastical fluctuation is at present not clear. Mass spectroscopy, in any case, should be postponed until a better knowledge of the instrument's response at high energies and data from longer observation periods become available.

 Turning to some illustrative examples of simulated sources, we show in Fig.~\ref{spectra} the situation of a model $A$ - type source, at a position (l,b)=(0,25) in galactic coordinates, with a mass $m_\chi=150$~GeV  and a flux $\phi=2\times 10^{-3}$ ph m$^{-2}$ s$^{-1}$. At this location, we expect a moderate diffuse background contribution;  it was chosen such as to demonstrate that a source corresponding to EGRET's faintest detected source
is easily identified. In the same figure, we also present  a source with $m_\chi=1$~TeV and $\phi = 2\times 10^{-2}$ ph m$^{-2}$ s$^{-1}$. Here, we have picked a position in the galactic plane, at (l,b)=(50,0), to show
that even in presence of an intense diffuse gamma-ray background, a cutoff at 1 TeV is well
within the reach of our analysis -- at least for
the brightest sources in our realizations. In the right panel of Fig.~\ref{spectra}, finally, we show an example spectrum for a bright ($\phi=2\times 10^{-2}$ ph m$^{-2}$ s$^{-1}$) model $B$ - type DM source with $m_\chi=150$~GeV. The source was simulated at galactic coordinates (l,b)=(0,50), where background counts are negligible;  the superimposed fit was computed with the usual Eq.~(\ref{eq:param}), i.e.~without taking into account the leptonic contribution.
At high energies, one clearly notices a photon excess due to the leptonic component; however, the photon counts are nonetheless rather small and therefore fluctuations play a major role.
As observation time increases, a leptonic component as in model $B$ would be recognizable as such even for fainter sources; for the brighter sources, on the other hand, we expect the characteristic, sharply pronounced cutoff to become ever more visible. For a more detailed discussion, a dedicated analysis would be necessary; we leave this interesting issue open for future studies. 

To conclude this Section, let us briefly return to the question of how to discriminate a DM from a mere astrophysical origin of gamma-ray point sources, which is certainly mostly based on the detection of the high energy cutoff at the mass of the DM particle. The GLAST LAT is expected to observe more than a hundred of Active Galactic Nuclei (AGNs) in the period we selected for this analysis (2 months), some of which would, due to EBL attenuation \cite{fazioagn}, feature a spectral signature similar to that of DM annihilations. However, a population of DM sources would exhibit distinctive features that should allow to discrminate the from AGNs, such as the low energy spectral index of $\sim 1.5$ and the absence of variability. Furthermore, a population of DM sources should appear as a distinctive excess in the Fazio-Stecker plot (see for example \cite{fazioagn}) that can be obtained from the LAT AGN catalog.

\section{Discussion and Conclusions}
\label{conclu}

The upcoming space telescope GLAST will perform full-sky observations 
of the gamma-ray sky, at energies that are of particular interest for 
indirect DM studies. In this paper, we have performed a 2 months 
simulation of the gamma-ray sky as observed by
the LAT instrument on board the satellite, and, assuming a gamma-ray spectrum typical for DM annihilations,
we have estimated the minimum flux, at all galactic latitudes and longitudes, 
required to {\it detect} the source at $5 \sigma$ above the background, and to
{\it identify} it, i.e. to discriminate it against ordinary astrophysical 
sources through the analysis of its spectral features. 

We have used this information 
to produce the sensitivity maps in Figs.~\ref{sensmap} and \ref{cutmap}, which represent the 
main result of this paper. The first one is a sensitivity map for the {\it detection}
of point sources of DM annihilation, the second  a sensitivity map
for the {\it identification} of its DM origin, and determination of the mass 
thanks to a fit of the high-energy cut-off. It should actually be possible to 
identify the DM origin of a population of gamma-ray sources also in absence 
of a precise determination of the high-energy cut-off, thanks to the distinctive 
value of the spectral slope in the low-energy tail, as discussed in 
Section II, the presence of an excess in the Fazio-Stecker plot, and absence of 
variability, as argued in Section VII. However, the precise determination of 
the high-energy cut-off would provide the opportunity of making an even stronger 
case for the 'exotic' origin of the gamma-ray signal, since astrophysical sources 
can hardly mimic DM spectral features (see also the recent paper by Baltz 
et al.~\cite{Baltz:2006sv}).

These maps represent a powerful tool to rapidly, and efficiently, investigate the detectability 
of any population of point sources of DM annihilation, such as small scale 
clumps~\cite{Diemand:2006ik,Koushiappas:2006qq,Pieri:2005pg,Pieri:2003cq,Koushiappas:2003bn,Bergstrom:1998zs},
compact DM structures like Spikes ~\cite{Gondolo:1999ef, Bertone:2002je, Ullio:2001fb, Bertone:2005hw,Bertone:2005xv} or Crests~\cite{Merritt:2006mt}, and DM mini-spikes
~\cite{Bertone:2005xz,Zhao:2005zr}. All these objects would appear as pointlike 
sources with GLAST, if their angular size is smaller than the angular resolution 
of the instrument, i.e. $\sim 0.1^\circ$. Objects of physical size $R_p$ would thus appear 
pointlike when at distance $D \gtrsim 5.7 \times 10^2 R_p$, a constraint often
satisfied by the objects listed above. 
   
In order to 
prove the effectiveness of the method, we have applied our results to the 
mini-spikes scenario discussed in Ref.~\cite{Bertone:2005xz}, consisting 
of a population of $\sim 100$ DM overdensities, dubbed
mini-spikes, around Intermediate Mass Black Holes. We were thereby able to show that 
a large number of these objects can be detected and identified with GLAST, if
they exist, while null searches would place extremely stringent constraints on the
whole scenario. Although a set of fiducial astrophysical parameters 
was chosen for the mini-spike scenario, in order to perform this study, the
analysis can easily, and rapidly, be generalized to any other set of parameters.

Finally, we have shown that, in case of detection, GLAST can
provide useful information on the nature of the annihilating DM particles: first, with a
relatively accurate determination of their mass, and second, with the 
identification of some specific spectral features, like a possible hard spectrum 
contribution near the cut-off at the DM mass, due to the annihilation to
charged leptons. 

\mbox{ }\vspace{1cm}

\section*{Acknowledgements}
We thank Andrew Zentner for making available
the numerical realizations of Intermediate Mass Black Holes in the Milky 
Way halo, on which our estimates are based. GB is supported by the Helmholtz 
Association of National Research Centres, under project VH-NG-006.


\end{document}